\documentclass[aps,samsmath,amssymb,showpacs,amsfonts,prd,nofootinbib]{revtex4-1}

\usepackage{graphicx}
\usepackage{dcolumn}
\usepackage{bm}
\usepackage{amsmath,graphics,color}
\usepackage{color,psfrag}
\usepackage{mathrsfs}

\newcommand{\beq}{\begin{equation}}
\newcommand{\eeq}{\end{equation}}
\newcommand{\bea}{\begin{eqnarray}}
\newcommand{\eea}{\end{eqnarray}}
\newcommand{\n}{{(n)}}
\def\half{{\textstyle{\frac{1}{2}}}}

\def\d{\partial}

\def\d{\delta}

\def\phi{\varphi}

\begin{document}

\title{Higher Curvature Gravity and the Holographic fluid dual to flat spacetime}

\author{Goffredo Chirco}
\email{chirco@sissa.it}
\author{Christopher Eling}
\email{cteling@sissa.it}
\author{Stefano Liberati}
\email{liberati@sissa.it}
\affiliation{SISSA, Via Bonomea 265, 34136 Trieste, Italy and INFN Sezione di Trieste, Via Valerio 2, 34127 Trieste, Italy}

\date{\today}

\begin{abstract}
Recent works have demonstrated that one can construct a $(d+2)$ dimensional solution of the vacuum Einstein equations that is dual to a $(d+1)$ dimensional fluid satisfying the incompressible Navier-Stokes equations. In one important example, the fluid lives on a fixed timelike surface in the flat Rindler spacetime associated with an accelerated observer.  In this paper, we show that the shear viscosity to entropy density ratio of the fluid takes the universal value $1/4\pi$ in a wide class of higher curvature generalizations to Einstein gravity. Unlike the fluid dual to asymptotically anti-de Sitter spacetimes, here the choice of gravitational dynamics only affects the second order transport coefficients. We explicitly calculate these in five-dimensional Einstein-Gauss-Bonnet gravity and discuss the implications of our results.
\end{abstract}
\maketitle

\section{Introduction}

In the past few years there has been increasing interest in the holographic duality relating fluid dynamics and gravity.  The link was first introduced in the 1970's with the development of the membrane paradigm approach \cite{membrane}, where the dissipative black hole horizon dynamics is recognized to closely resemble that of a viscous fluid. The connection has been made more concrete in the context of the AdS/CFT correspondence, where quantum gravity in an asymptotically anti-de Sitter (AdS) spacetime is shown to be dual to a certain gauge theory in flat spacetime in one lower dimension \cite{Maldacena:1997re}. The gauge theory can be thought of as living on the timelike AdS boundary, in which the bulk spacetime is holographically encoded. As a consequence, the relativistic hydrodynamics of the gauge theory can be effectively described by the long time, long wavelength dynamics of a black hole in AdS \cite{Bhattacharyya:2008jc}. The relativistic Navier-Stokes equations turn out to be equivalent to the subset of the General Relativity (GR) field equations called the momentum constraints, which constrain ``initial" data on the timelike AdS boundary. Moreover, the incompressible Navier-Stokes equations describing ordinary, everyday fluids can be obtained by taking a particular non-relativistic limit of these results \cite{Bhattacharyya:2008kq}.

A key step toward a deeper understanding of the fluid/gravity correspondence can be found in the question of whether an asymptotically AdS spacetime is actually a required ingredient. Indeed, there have been a number of hints indicating that this is not the case. For example, the momentum constraint equations are not affected by the value of the cosmological constant, which suggests that the full asymptotic structure of the spacetime is un-important. Secondly, in cases where one can perform a hydrodynamic (long time, long wavelength) expansion of the equations describing the horizon dynamics, one finds they also have the form of Navier-Stokes equations \cite{nonrel, rel,pad}. Interestingly, one example where such an expansion exists is for a Rindler acceleration horizon in flat spacetime \cite{Eling:2008af, nonrel}.

In a recent paper \cite{Bredberg:2010ky}, a novel formalism was introduced to describe a holographic fluid theory defined on an arbitrary timelike surface in a general spacetime with a causal horizon. On this surface, one fixes the boundary condition that the induced metric is flat, and in the spirit of the Wilsonian approach to the renormalization, the asymptotic physics outside this surface plays no role. Moving this surface between the horizon and the asymptotic boundary can be thought of as a renormalization group flow between a boundary fluid and a horizon fluid. In \cite{Bredberg:2011jq} the authors considered the specific case of perturbations about a Rindler metric, taking the timelike surface to be one of the family of hyperbolas associated with the worldlines of an accelerated observer.  Working in the non-relativistic hydrodynamic expansion, the authors presented a geometry that is a solution to the Einstein equations if the data on surfaces of $r_c$ satisfy the incompressible Navier-Stokes equations.  Alternatively, one can consider the physically inequivalent near-horizon expansion in small $r_c$ and obtain the same results.

Beyond the connection between the classical Navier-Stokes equations and a classical geometry, these works actually suggested the possibility of an underlying holographic duality relating a theory on fixed $r_c$ to the interior bulk of the Rindler spacetime. A first step toward a detailed study of the behavior of this dual system was taken in \cite{Compere:2011dx}, with the introduction of an algorithm for constructing the geometry and the explicit expression for the viscous transport coefficients to second order in the hydrodynamic expansion.

In this paper, our main goal is to probe further the dual theory by asking what effect higher curvature terms in the dual gravitational theory have on the transport coefficients of the fluid dual to the Rindler geometry. In the AdS/CFT correspondence, such terms are associated with quantum corrections or other deformations, which modify the values of the transport coefficients. Remarkably, we show here that the shear viscosity of the Rindler fluid is not modified if higher curvature terms are introduced. Equivalently, at lowest orders in the non-relativistic expansion, the dual metric solution has the property of being a solution to GR and to any higher curvature theory of gravity. The first place the higher curvature corrections appear is in the second order transport coefficients of the fluid. Working in the case where the higher curvature theory is Einstein-Gauss-Bonnet gravity, we calculate some of these coefficients.

The plan of this paper is as follows. In Section II, we describe the general construction of the solutions developed in \cite{Bredberg:2010ky,Bredberg:2011jq,Compere:2011dx}. In Section III,  we explicitly show that the shear viscosity of the dual theory is unchanged when generic higher curvature terms are added into the gravitational action and we discuss the differences between this calculation and previous literature on the AdS/CFT correspondence. Section IV is devoted to the calculation of the second order transport coefficients in Einstein-Gauss-Bonnet. We then conclude with a discussion of the implications of these results and their possible connection to approaches using the local Rindler geometry as a tool for a thermodynamical derivation of gravitational dynamics.

\section{General Setup}

We want to construct a Lorentzian geometry that acts as the holographic dual description of a fluid flow in $d+1$ dimensions. Based on the holographic principle, we expect the fluid is defined on a $d+1$ dimensional timelike surface $S_c$ embedded in a $d+2$ dimensional bulk spacetime. We choose the timelike surface to be defined by fixed bulk radial coordinate, $r=r_c$. We also specialize to the case where the fluid moves on a flat background. In this case, the induced metric on $S_c$ should be flat as well, e.g.
\beq
\gamma_{\mu \nu} dx^\mu dx^\nu = -\Phi(r_c) dt^2 + e^{2 \Psi(r_c)} dx_i dx^i  \label{induced},
\eeq
where $\Phi$ and $\Psi$ are some functions of $r$.  We use the notation that coordinates on the hypersurface $S_c$ are $x^\mu = (t, x^i)$, where $i=1...d$. The $(d+2)$ dimensional bulk coordinates are defined with the notation $x^A = (t, x^i, r)$.  The final requirement is that the bulk spacetime must contain a regular, stationary causal horizon.  The bulk spacetime therefore has a timelike Killing vector field, which becomes null on the horizon. The full bulk metric therefore has the general form \cite{Bredberg:2010ky},
\beq
ds^2 = -\Phi(r) dt^2 + 2 dt dr + e^{2 \Psi(r)} dx_i dx^i,  \label{generalmetric}.
\eeq
where at some radius $r=r_h$ there is a horizon where $\Phi(r) = 0$ and the timelike Killing vector $\chi^A = (\partial_t)^A$ becomes null.
If one considers quantum field theory on the background (\ref{generalmetric}), one finds equilibrium thermal states associated with the presence of the horizon. For example, one can compute the Hawking temperature (in units where $\hbar=c=1$)
\begin{align}
T_H = \frac{\kappa}{2\pi} = \frac{\Phi'(r_h)}{4\pi},
\end{align}
where the surface gravity $\kappa$ can be defined via $\chi^B \nabla_B \chi^A = \kappa \chi^A$. Dividing by the redshift factor at $r_c$, $\sqrt{-g_{tt}} = \sqrt{\Phi(r_c)}$ yields the local Tolman temperature
\begin{align}
T_{loc} = \frac{\Phi'(r_h)}{4\pi \sqrt{\Phi(r_c)}}.
\end{align}
There is also an associated Bekenstein-Hawking entropy proportional to the cross-sectional area of the horizon
\begin{align}
S_{BH} = 4\pi e^{d \Psi(r_h)},
\end{align}
where here and throughout we use units such that $16\pi G= 1$. We want to identify these thermodynamical properties with the thermodynamical properties of the dual fluid in $d+1$ dimensions. Therefore, the general metric can be thought of as the dual geometrical description of an equilibrium thermal state associated with some lower dimensional theory defined on the surface $r=r_c$.

The metric (\ref{generalmetric}) can describe many different black hole solutions. Here we will focus on the special case of a region of flat $(d+2)$ dimensional Minkowski spacetime in ``ingoing Rindler" coordinates
\beq
ds^2 = -r dt^2 + 2 dt dr + dx_i dx^i \label{Rindler},
\eeq
where in terms of the above parametrization, $\Phi(r) = r$ and $\Psi(r) =0$. The null surface $r=0$ acts as a horizon to accelerated observers, whose worldlines correspond to surfaces of constant $r=r_c$.

Although the Rindler metric is just a patch of flat spacetime, the associated quantum field theory on this background has many of the same properties as a black hole solution. In particular,  surfaces of $r=r_c$ have a local Unruh temperature
\beq T = \frac{1}{4\pi \sqrt{r_c}}. \eeq
Strictly speaking, a Rindler horizon does not have a Bekenstein-Hawking entropy density. However, one can assign the Rindler horizon this entropy based on the holographic principle, or, more concretely, take the entropy to be the thermal entanglement entropy of the quantum fields in Rindler wedge \cite{entanglement}. This statistical entropy scales like an area, but is a UV divergent quantity. If a Planck scale cutoff is chosen appropriately, the entanglement entropy agrees with the Bekenstein-Hawking formula, i.e.
\beq s= 4\pi. \eeq
Given the existence of an equilibrium Unruh temperature and a Bekenstein-Hawking entropy density, the metric (\ref{Rindler}) can be thought of as a dual geometrical description of a perfect fluid in one lower dimension. This duality can be formalized by considering the Brown-York stress energy tensor \cite{Brown:1992br}, which in GR takes the form,
\beq
T^{BY}_{\mu \nu} = 2(K \gamma_{\mu \nu} - K_{\mu \nu}),  \label{BY0}
\eeq
where $K_{\mu \nu} = \frac{1}{2} \mathcal{L}_N \gamma_{\mu \nu}$ and $\mathcal{L}_N$ is the Lie derivative along the normal to the slice $N^A$.  One can show that $T^{BY}_{\mu \nu}$ (and its generalization for higher curvature gravity) is indeed equivalent to the stress energy tensor of the perfect fluid with a rest frame energy density $\rho$ and pressure $P$. In this case
\begin{align}
\rho = 0, \quad p = \frac{1}{\sqrt{r_c}} \label{rhoandP}.
\end{align}

\section{Equivalence of viscous hydrodynamics in Einstein and higher curvature gravities} \label{3}

\subsection{The seed metric}

In this section we will argue that the first order viscous hydrodynamics of the fluid defined on $S_c$ is independent of whether the dual gravitational theory is Einstein or some higher curvature generalization. In order to study the hydrodynamics of this fluid, we must perturb the background Rindler geometry. To start, we review the formalism for perturbing the Rindler metric developed in \cite{Compere:2011dx}. The first step is to make a set of coordinate transformations to obtain a new metric (or class of metrics). These transformations should keep the induced metric at $r_c$ flat. The transformed metric should also preserve a perfect fluid form of the stress energy tensor associated to the slice, as well as the time-like Killing vector and the homogeneity in the $x^i$ direction. It was shown in \cite{Compere:2011dx} that these set of conditions uniquely identify the two diffeomorphisms, namely a boost and the translation.

The boost of the metric takes the form,
\begin{align}
\sqrt{r_c} t \rightarrow \sqrt{r_c} t - \gamma \beta_i x^i, \quad  x^i \rightarrow x^i - \gamma \beta^i \sqrt{r_c} t + (\gamma-1) \frac{\beta_i \beta_j}{\beta^2} x^j,
\end{align}
where $\gamma = (1-\beta^2)^{-1/2}$ and $\beta_i = r_c^{-1/2} v_i$ is the boost parameter.
The linear shift of the radial coordinate and re-scaling of $t$, which moves the horizon from $r=0$ to an $r=r_h < r_c$, is instead
\begin{align}
r \rightarrow r-r_h, \quad  t \rightarrow (1-r_h/r_c)^{-1/2} t.
\end{align}
The resulting metric for the flat spacetime is
\begin{align}
d s^2 &= \frac{dt^2}{1-v^2/r_c}\left(v^2 - \frac{r-r_h}{1-r_h/r_c}\right)+\frac{2\gamma}{\sqrt{1-r_h/r_c}}dt dr - \frac{2\gamma v_i}{r_c \sqrt{1-r_h/r_c}}d x^i dr \nonumber \\[1ex]
&\quad + \frac{2v_i}{1-v^2/r_c}\left( \frac{r-r_c}{r_c-r_h}\right) d x^i dt + \left(\delta_{ij}- \frac{v_iv_j}{r_c^2 (1-v^2/r_c)} \left(\frac{r-r_c}{1-r_h/r_c}\right)\right) dx^i d x^j \label{genmetric}.
\end{align}

We now want to investigate the hydrodynamic system dual to the above metric. To do that, we need to consider the dynamics of the metric perturbations within a hydrodynamic limit. One can perturb (\ref{genmetric}) by promoting the spatial velocity and horizon radius to be functions of space and time: $v^i(t, x^i)$ and $r_h(t, x^i)$.  Now the metric is no longer flat and no longer a solution of the vacuum Einstein equation. However, one can introduce a particular non-relativistic hydrodynamical expansion \cite{Fouxon:2008tb, Bhattacharyya:2008kq} in terms of a small parameter $\epsilon$,
\begin{align}
v^i \sim \epsilon v^i (\epsilon x^i , \epsilon^2 t) \quad P \sim \epsilon^2 P(\epsilon x^i , \epsilon^2 t), \label{scaling}
\end{align}
where the \textit{non-relativistic} pressure $P(t,x^i)$ is defined in the following way as a small perturbation of the horizon radius, \footnote{Note that the $\epsilon$ expansion is performed in such a way that at zeroth order $v^i = r_h=0$ so that the standard Rindler metric (\ref{Rindler}) is recovered. Also, there is no scaling of bulk radial derivatives.}
\beq
r_h = 0 + 2 P + O(\epsilon^4).
\eeq

Using (\ref{scaling}) one scales down the amplitudes ($\epsilon$ can be thought of as the inverse of the speed of light), while at the same time scaling to large times $t$ and spatial distances $x^i$. This corresponds to looking at small perturbations in the hydrodynamic limit.

Expanding the metric (\ref{genmetric}) out to $O(\epsilon^2)$ in this manner yields the ``seed metric" solution originally found by Bredberg, Keeler, Lysov and Strominger in \cite{Bredberg:2011jq},
\begin{align}
d s^2 &= -r dt^2+2dt dr+d x_i d x^i \nonumber\\
&\quad -2\left(1-\frac{r}{r_c}\right)v_i d x^i dt -\frac{2v_i}{r_c} d x^i d r \nonumber\\
&\quad +\left(1-\frac{r}{r_c}\right)\Big[(v^2+2P) dt^2+\frac{v_iv_j}{r_c} d x^i d x^j\Big]+\left(\frac{v^2}{r_c}+\frac{2P}{r_c}\right)d t dr.
\label{seedmetric}
\end{align}
The seed metric is the unique singularity-free solution to the vacuum Einstein equations up to $O(\epsilon^3)$, provided $\partial_i v^i = 0$. As required, the induced metric on the slice $r=r_c$ is flat.

In GR, the momentum constraint equations on the surface $S_c$ can be expressed in terms of the Brown-York stress tensor
\beq R_{\mu A} N^A = \partial^\nu T^{BY}_{\mu \nu} = 0. \label{2ndconstr}
\eeq
At second and third order in $\epsilon$, momentum constraint equations are
\beq R^{(2,3)}_{\mu A} N^A = r_c^{-1/2} R^{(2,3)}_{t\mu} + r_c^{1/2} R^{(2,3)}_{r \mu} = 0, \label{3rdconstr}
\eeq
while the Brown-York stress-tensor for the seed metric is given by \cite{Bredberg:2011jq}
\begin{align}
\label{seedstress}
T^{BY}_{\mu \nu} d x^\mu d x^\nu &= \frac{d \vec{x}^2}{\sqrt{r_c}}
- \frac{2 v_i}{\sqrt{r_c}}\, d x^i dt
+\frac{v^2}{\sqrt{r_c}}\,d t^2 + r_c^{-3/2}\Big[P\delta_{ij}+v_iv_j-2 r_c \partial_i v_j\Big] d x^i d x^j +O(\epsilon^3)\, .
\end{align}
Then, at second order, using the expression in (\ref{seedstress}), the momentum constraint equations (\ref{2ndconstr}) reduce to the incompressibility condition $\partial_i v^i =0$ we discussed above.  At third order one finds the Navier-Stokes equations with a particular kinematic viscosity
\beq
\partial_t v_i + v^j \partial_j v_i + \partial_i P - r_c \partial^2 v_i = 0. \label{NS}
\eeq
Therefore, imposing the the incompressible Navier-Stokes equations on the fluid variables guarantees the dual metric is a solution to the field equations.

Noticeably, these results can be obtained as a non-relativistic expansion of a relativistic viscous fluid stress tensor. To see this, we work in the relativistic hydrodynamic expansion in derivatives of the fluid velocity and pressure: $\partial u$ and $\partial p$. Then, at first order, the relativistic viscous fluid stress tensor has the form,
\beq
T^{\mathrm{fluid}}_{\mu \nu} = \rho u_\mu u_\nu + p h_{\mu \nu} - 2 \eta K_{\mu \nu} - \xi h_{\mu \nu} (\partial_\lambda u^\lambda). \label{viscousrelT}
\eeq
Here $h_{\mu \nu} = \gamma_{\mu \nu} + u_\mu u_\nu$, while $K_{\mu \nu} = h^\lambda_\mu h^\sigma_\nu \partial_{(\lambda} u_{\sigma)}$ is the fluid shear, $\eta$ the shear viscosity, and $\xi$ the bulk viscosity.

The viscous terms above are written in the Landau or transverse frame \cite{Landau}, which can be defined as a condition on the first order part of the stress tensor
\beq  T^{\mathrm{fluid}~ (1)}_{\mu \sigma} u^\sigma = 0. \label{Landau}\eeq
This frame is constructed so that the viscous fluid velocity is defined as the velocity of energy transport.  The seed stress tensor in (\ref{seedstress}) follows from the $\epsilon$ expansion of (\ref{viscousrelT}), if we identify
\begin{align}
u^\mu = \frac{1}{\sqrt{r_c-v^2}}(r_c, v^i) , \quad \rho = 0+ O(\epsilon^3), \quad p = \frac{1}{\sqrt{r_c}} + \frac{P}{r_c^{3/2}}, \quad \eta = 1. \label{fluiddata}
\end{align}
This is consistent with the earlier equilibrium calculation of $\rho$ and $p$ in (\ref{rhoandP}). Note also that the bulk viscosity term in (\ref{viscousrelT}) actually drops out and bulk viscosity is not an independent transport coefficient. This is due to the fact that at viscous order we can impose the ideal order equation $\partial_\mu u^\mu = 0$, which follows from $\rho=0$ and continuity.

\subsection{Higher Curvature Gravity}

Now we want to study how the hydrodynamics of the fluid is modified when the gravity theory is not GR, but instead some theory with higher curvature terms. The first question is whether we need a new, modified seed metric in a higher curvature theory of gravity. Interestingly, we can show that the seed metric (\ref{seedmetric}) and its $O(\epsilon^3)$ correction is a solution to a wide class of higher curvature gravity theories at lowest orders in the $\epsilon$ expansion.

We start by noting that the flat, equilibrium Rindler metric at zeroth order is a vacuum solution to both Einstein and higher curvature gravity theories. The higher curvature terms could be thought of as modified gravity theories in their own right or they can be seen as quantum corrections to Einstein gravity in an effective field theory picture. Here we will not consider exotic theories involving inverse powers of curvature invariants.

As a first example of a higher curvature theory we consider Einstein-Gauss-Bonnet gravity (in the absence of a cosmological constant), defined by the action
\beq I_{GB} = \int d^{d+2} x \sqrt{-g} \left[R + \alpha \left (R^2 - 4 R_{CD} R^{CD} + R_{CDEF} R^{CDEF}\right) \right], \label{GBaction}
\eeq
where $\alpha$ is the Gauss-Bonnet coupling constant. We consider $d \geq 3$ since for $d<3$ the Gauss-Bonnet term is topological and does not affect the field equations. The interest in looking at a Gauss-Bonnet term is twofold. Such a term arises in the low energy limit of string theories. Secondly, Einstein-Gauss-Bonnet gravity is notable because even though the action is higher order in the curvature, for the unique combination of curvature invariants in the second term of (\ref{GBaction}), the field equations remain second order in derivatives of the metric.

Varying this action with respect to the metric yields the field equations,
\beq
G_{AB} + 2 \alpha H_{AB} = 0 \label{EGBfieldeqn},
\eeq
where the Lovelock tensor $H_{AB}$ is
\begin{align}
H_{AB} = R R_{AB} - 2 R_{AC} R^C_B - 2 R^{CD} R_{ACBD} + R_A{}^{CDE} R_{BCDE}- \frac{1}{4} g_{AB} \left(R^2 - 4 R_{CD} R^{CD} + R_{CDEF} R^{CDEF}\right) \label{Lovelock}.
\end{align}
Now, using the seed metric, the first non-zero components of the Riemann tensor $R_{ABC}{}^D$ are at $O(\epsilon^2)$. If we examine the Lovelock tensor, (\ref{Lovelock}), it is clear that the first contributions from the Gauss-Bonnet terms can only appear at $O(\epsilon^4)$ at the lowest. A similar conclusion obviously holds for Lovelock gravities \cite{Lovelock:1971yv}, which are the extension of the action (\ref{GBaction}) including contributions with higher powers of the curvature but still yielding 2nd order field equations.

The field equations of other higher curvature theories of gravity generally involve covariant derivatives of the Riemann tensor and its contractions. These are no longer second order in metric derivatives. At second order in the curvature the gravitational action has the form

\beq
I = \int d^{d+2} x \sqrt{-g} \left(R + \beta_1 R^2 + \beta_2 R_{AB} R^{AB} + \beta_3 R_{ABCD} R^{ABCD}\right) \label{R2action}.
\eeq
The field equations can be expressed in the form $G_{AB} = S^{eff}_{AB}$, where
\begin{align}
S^{eff}_{AB} &= \beta_1 \left(R R_{AB} - \nabla_A \nabla_B R + g_{AB} (\Box R - \frac{1}{2} R^2)\right) + \nonumber \\ &
\beta_2 \left(g_{AB} R_{CD} R^{CD} + 4 \nabla_C \nabla_B R^C_A - 2 \Box R_{\mu \nu} - g_{AB} \Box R - 4 R^C_A R_{CB} \right) \nonumber \\ & \quad +\beta_3 \left(g_{AB} R_{ABCD} R^{ABCD} - 4 R_{ACDE} R_{B}{}^{CDE} - 8 \Box R_{AB} +  4 \nabla_B \nabla_A R + 8 R^C_A R_{CB} - 8 R^{CD} R_{ACBD} \right) \label{Seff}.
\end{align}
Let's consider the possible terms that can appear at the lowest orders in $\epsilon$. First, the second covariant derivative terms of $R$ could in principle contribute $\beta_i$ corrections at $O(\epsilon^2)$. However, the Ricci scalar $R = g^{AB} R_{AB}$ can be expanded out as follows,
\beq R= g^{tt} R_{tt} + 2 g^{rt} R_{tr} + 2 g^{ti} R_{ti} + 2 g^{ri} R_{ri}+ g^{rr} R_{rr} + g^{ij} R_{ij}. \label{Rexp}\eeq
Before imposing incompressibility, one can show that the only non-zero component of $R_{AB}$ at $O(\epsilon^2)$ is
\beq R_{tt} = \half \partial_i v^i. \eeq
However, for the background Rindler metric (\ref{Rindler}), the zeroth order $g^{tt}{}_{(0)}$ is zero, so the Ricci scalar $R$ is in fact higher order. Since one cannot form a scalar constructed from $v^i$, $P$, $\partial_t$, and $\partial_i$ with odd powers of $\epsilon$, we expect $R$ is of $O(\epsilon^4)$. For instance, the spatial vector $R_{ti}$ is $O(\epsilon^3)$, but this multiplies $g^{ti}$, which is $O(\epsilon)$. Therefore, $R$ is $O(\epsilon^4)$ and its covariant derivatives are of the same order or higher.

The remaining terms of interest are the $\Box R_{AB}$ and $ \nabla_C \nabla_B R^C_A$ terms proportional to $\beta_2$ and $\beta_3$. We know that $R_{AB}$ a priori has non-zero components at $O(\epsilon^2)$ and $O(\epsilon^3)$.  The question is whether the radial derivatives and background connection for the Rindler metric (\ref{Rindler}) allow the above two terms to also contribute at these orders in $\epsilon$ thereby affecting the hydrodynamics at these orders. This we checked with an explicit calculation. The result is again negative.

Thus, as a general principle, higher curvature corrections to the Einstein equations come in at $O(\epsilon^4)$, at least when we perturb the fluid dual to the flat Rindler spacetime geometry. Terms of even higher order in the curvature (schematically $\sim R^n$, where $n > 2$) will typically appear at even higher orders. This includes the often studied case of $f(R)$ theories, when $f$ can be expanded around the Hilbert term: $f = R + R^2 + R^3 + \cdots$.

As a result, the solution to the higher curvature theories at the lowest orders $O(\epsilon^2)$ and $O(\epsilon^3)$ is the same as the GR solution found previously \cite{Bredberg:2011jq, Compere:2011dx}.  Since all the higher curvature quantities vanish at the lowest orders, this solution has the property of being approximately \textit{strongly universal} \cite{Coley:2008th} . The explicit solution at $O(\epsilon^3)$ can be constructed from the algorithm for Einstein gravity given in \cite{Compere:2011dx}, which we will expand upon and generalize to Einstein-Gauss-Bonnet in the next section.  At the present, we note that the equivalence of the solutions to $O(\epsilon^4)$ implies that the 1st order viscous hydrodynamics of the dual fluid is the same both in Einstein gravity and its higher curvature generalizations.
In particular, the incompressible Navier-Stokes equations (\ref{NS}) are the same in any theory, with the kinematic viscosity fixed to be $r_c$. Furthermore, as Comp\`ere, et. al. pointed out, it is clear that the non-relativistic $\epsilon$ expansion is capturing the non-relativistic limit of a relativistic fluid theory whose full structure is unknown. Nevertheless, the $\epsilon$ expansion seems to be able to capture some of the transport properties of this fluid theory. In particular, the shear viscosity of the relativistic fluid, $\eta$, is apparently fixed to be $1$ (or $(16\pi G)^{-1}$ if we restore the gravitational constant).

One may worry about using the non-relativistic limit to draw conclusions about the properties of the relativistic parent fluid. However, we can show that our analysis of higher curvature terms can be extended to the relativistic hydrodynamics. The first step is write the metric (\ref{genmetric}) in a manifestly boost covariant form. This metric turns out to be
\begin{align}
ds^2 = -(1+p^2(r-r_c)) u_\mu u_\nu dx^\mu dx^\nu - 2 p u_\mu dx^\mu dr + h_{\mu \nu} dx^\mu dx^\nu. \label{relativ}
\end{align}
In this line element we have replaced $r_h$ with the relativistic pressure $p$ using the general formula
\beq
p = \frac{1}{\sqrt{r_c-r_h}}.
\eeq
Expanding $u^\mu$ and $p$ in terms of $v^i$ and $P$ using (\ref{fluiddata}), the metric (\ref{relativ}) reproduces the seed metric up to $O(\epsilon^2)$. In addition, if we compute the Brown-York stress tensor at $r=r_c$ for this metric (\ref{relativ}), we find directly
\begin{align}
T_{\mu \nu} dx^\mu dx^\nu = p h_{\mu \nu} dx^\mu dx^\nu \label{relstress1},
\end{align}
which is the ideal part of (\ref{viscousrelT}) with $\rho=0$.

To perturb in this case, we now treat $u^\mu(x^\mu)$ and $p(x^\mu)$, but leave $r_c$ fixed. The metric is no longer a solution to the vacuum Einstein equations, but one can expand and work order by order in derivatives of $u^\mu$ and $p$ as discussed earlier. This follows the standard approach used in the fluid-gravity correspondence \cite{Bhattacharyya:2008jc}.

We now have that (\ref{relativ}) is a zeroth order solution, i.e. $R_{AB} = 0 + O(\lambda)$, where the parameter $\lambda$ counts derivatives of $u^\mu$ and $p$. Therefore, $R_{ABC}{}^D \sim O(\lambda)$ and the curvature squared terms in (\ref{Seff}) must appear at $O(\lambda^2)$.  The other terms involve the covariant derivatives of the Ricci scalar and tensor. The generalization of (\ref{Rexp}) is
\beq
R^{(1)} = g^{rr} R_{rr}^{(1)}  + 2 g^{r\mu} R_{r\mu}^{(1)}  + g^{\mu \nu} R_{\mu \nu}^{(1)} .
\eeq
From (\ref{relativ}) we find
\begin{align}
\label{relRicci}
R_{rr}^{(1)} &= 0 \nonumber \\
R_{r \mu}^{(1)} &= 0 \nonumber \\
R_{\mu \nu}^{(1)} &= \partial_{(\mu} p u_{\nu)} + D p ~u_\mu u_\nu + \frac{1}{2} p (\partial_\lambda u^\lambda) u_\mu u_\nu + p u_{(\mu} a_{\nu)},
\end{align}
where we have defined $D = u^\mu \partial_\mu$ and $a_\mu = u^\lambda \partial_\lambda u_\mu$. Since $g^{\mu \nu} = h^{\mu \nu}$, which projects orthogonal to $u^\mu$, $R^{(1)} =0$.  Finally, the fact that the remaining terms $\Box R_{AB}$ and $ \nabla_C \nabla_B R^C_A$ are also of $O(\lambda^2)$ can be shown by explicit calculation as before.

Therefore, we conclude again that the higher curvature terms affect only the second order viscous hydrodynamics.  The equilbrium stress tensor will be given by Eqn. (\ref{relstress1}) in any higher curvature theory of gravity. This follows just from the fact that the zeroth order metric (\ref{relativ}) is a solution in any theory. Computing the $O(\lambda)$ corrections to (\ref{relativ}) and (\ref{relstress1}) confirms that $\eta = 1$ and the bulk viscosity is not a transport coefficient, but we will save the details for another paper \cite{WIP}.

In higher curvature theories, the entropy is given by the Wald formula \cite{Wald:1993nt}. In general,  Bekenstein-Hawking area entropy will be modified by the higher curvature terms, leading to an expression that can depend on both the intrinsic and extrinsic geometry of horizon. However, since we are working with a Rindler horizon in flat spacetime, all these corrections vanish and the equilibrium entropy density $s$ remains $4\pi$. The ratio $\eta/s =1/4\pi$ was first derived in the context of the AdS/CFT correspondence \cite{Policastro:2001yc}. It was shown that the ratio goes to this value for any infinitely strongly coupled holographic gauge theory fluid with an Einstein gravity dual \cite{Buchel:2003tz}. On the gauge theory side, the number of colors $N \rightarrow \infty$ and the 't Hooft coupling $\lambda \rightarrow \infty$.  This is essentially a classical limit; quantum corrections to the $\eta/s$ ratio at finite $N$ and $\lambda$, which can be calculated in specific string theory realizations \cite{Myers:2008yi}, correspond to specific higher derivative corrections to the dual gravitational theory. Another approach is to work outside the context of particular string theories and consider a generic higher curvature gravity action of the form given in (\ref{R2action}). In this case, it has been shown \cite{GBviscosity,Cai:2011xv} that ratio changes to
\beq
\frac{\eta}{s} = \frac{1}{4\pi}(1- 8 \beta_3). \label{ratioAdS}
\eeq
This result holds in five spacetime dimensions and to linear order in the $\beta_i$, which are effectively suppressed by powers of the Planck length. It is also important to note that while the ratio is unchanged when $\beta_3=0$, both $\eta$ and $s$ do depend on $\beta_{1,2}$. Finally, in the special case of Einstein-Gauss-Bonnet, (\ref{GBaction}), $\beta_3 = \alpha$. Given the nice properties of this theory (linked to the field equations remaining 2nd order in derivatives), one can work non-perturbatively and consider finite $\alpha$ corrections which allow the ratio to approach zero.

It is then remarkable that in the case of a flat Rindler background there is no higher curvature correction to the ratio or to the viscosity itself. The viscosity is protected against quantum corrections or other deformations to the dual theory. At a technical level, the difference is that the result (\ref{ratioAdS}) follows by considering perturbations around a background asymptotically AdS black brane solution in the higher curvature gravity theory. In Einstein-Gauss-Bonnet gravity with negative cosmological constant, this solution is \cite{Cai:2001dz}
\begin{align}
ds^2 = N^2 f(r) - \frac{1}{f(r)} dr^2 + r^2 dx_i dx^i
\end{align}
where $N$ is some constant and
\begin{align}
f(r) = \frac{r^2}{4\alpha} \left(1-\sqrt{1-8\alpha \left(1-\frac{r_h^4}{r^4}\right)}\right),
\end{align}
with $r_h$ the value of the horizon radius. In this solution, thermodynamic quantities such as the temperature and entropy density depend explicitly on $\alpha$, which translates into the calculations of the entropy and shear viscosity.

In contrast, in the Rindler case the metric does not depend on $\alpha$ and the Unruh temperature and entanglement entropy are kinematical quantities in the sense that they are independent of the underlying gravitational theory.  The shear viscosity seems to have the same behavior since it is also unaffected by the choice of gravitational dynamics. This is further evidence for the picture of $\eta/s= 1/4\pi$ as a kinematical property associated with entanglement in Rindler spacetime \cite{Chirco:2010xx}.

\section{Second order transport coefficients}

Now let's consider the hydrodynamic expansion at higher order in derivatives.  Here we expect the gravitational dynamics to affect the hydrodynamics of the dual fluid. To second order, $O(\lambda^2)$, the general stress tensor for a relativistic fluid with zero energy density (hence incompressible) has the form \cite{Compere:2011dx}
\begin{align}
T^{\mathrm{fluid}}_{\mu \nu} &= \rho u_\mu u_\nu + p h_{\mu \nu} - 2 \eta K_{\mu \nu} \nonumber \\ & \quad + c_1 K_\mu^\lambda K_{\lambda \nu} + c_2 K_{(\mu}^\lambda \Omega_{|\lambda|\nu)} + c_3 \Omega_\mu^{\,\,\,\lambda}\Omega_{\lambda \nu} +
c_4 P_\mu^\lambda P_\nu^\sigma D_\lambda D_\sigma \ln p \nonumber\\
& \quad  + c_5 \sigma_{\mu \nu}\,D\ln p + c_6 D^\perp_\mu \ln p \,D^\perp_\nu \ln p, \label{generalstress}
\end{align}
where $D= u^\mu \partial_\mu$, $D^\perp_\mu = P^\nu_\mu \partial_\nu$, and $\Omega_{\mu \nu} = P_\mu^\lambda P_\nu^\sigma\partial_{[\lambda}u_{\sigma]}$. There are also viscous corrections to the energy density $\rho$ at this order, which can be parameterized as
\begin{align}
\rho = b_1 K_{\mu \nu}K^{\mu \nu}+ b_2 \Omega_{\mu \nu}\Omega^{\mu \nu}+ b_3 D\ln p \,D\ln p + b_4 D^2 \ln p+b_5 D^\perp_\mu \ln p \,D^{\perp \mu}\ln p.
\end{align}
The $c_i$, $i=1..6$, and $b_j$, $j=1..5$, are the possible new transport coefficients. When one expands these expressions in powers of $\epsilon$, many of the second order transport coefficients appear at $O(\epsilon^4)$ in a general non-relativistic fluid stress tensor,
\begin{align}
T^{\mathrm{fluid}~(4)}_{\mu \nu}d x^\mu d x^\nu &=  r_c^{-3/2}\Big[v^2(v^2+P)-\eta r_c \sigma_{ij}v^i v^j + \frac{b_1 r_c^{3/2}}{2}\sigma_{ij}\sigma^{ij} + \frac{b_2 r_c^{3/2}}{2} \omega_{ij} \omega^{ij}
\Big]dt^2\nonumber \\[1ex]
&\quad + r_c^{-5/2}\Big[ v_iv_j (v^2+P) +2\eta r_c v_{(i}\partial_{j)}P + c_4 r_c^{3/2} \partial_i \partial_j P+\frac{c_1}{4}r_c^{3/2}\sigma_{ik}\sigma^{k}{}_{j}
\nonumber\\[1ex]
&\quad\qquad + \frac{c_3}{4}r_c^{3/2} \omega_{ik} \omega^{k}{}_{j}
-\frac{c_2}{4}r_c^{3/2}\sigma_{k(i}\omega_{j)}{}^{k} -2\eta r_c^2 v_{(i}\partial^2 v_{j)} -\eta r_c v_{(i}\partial_{j)}v^2 \nonumber\\[1ex]
&\quad\qquad
-\frac{r_c}{2}\eta \sigma_{ij}v^2  \Big] d x^i d x^j . \label{T4}
\end{align}
Here $\sigma_{ij} = 2 \partial_{(i} v_{j)}$ and $\omega_{ij} = 2 \partial_{[i} v_{j]}$. Only $c_5$, $c_6$ and $b_3$, $b_4$, and $b_5$ are absent at this order in the $\epsilon$ expansion.

We argued that $O(\epsilon^4)$ is the first to receive corrections from any higher curvature terms in the gravity theory. In the next section, we will solve for the fourth order (non-relativistic) metric in five dimensional Einstein-Gauss-Bonnet gravity. With this result in hand, we will use the corresponding Brown-York stress tensor to read-off various second order transport coefficients for the dual fluid.

\subsection{Constructing the Einstein-Gauss-Bonnet solution}

We first outline the construction due to \cite{Compere:2011dx}, where one starts with the metric solution at $O(\epsilon^{n-1})$. In practice, the first $n$ is $3$ , i.e. one starts the process with the seed metric solution (\ref{seedmetric}). We then want to add to the metric a new piece $g^\n_{AB}$ that solves the field equations to $O(\epsilon^{n+1})$. Since radial derivatives carry no powers of $\epsilon$, the addition of $g^\n_{AB}$ produces a change in the bulk curvature tensors at the same order. This is effectively a perturbation around the zeroth order background Rindler metric (\ref{Rindler}). We work in the gauge where
\beq g^\n_{rA} =0, \eeq
for all the contributions with $n\ge 3$. With this choice, we find that changes in the Einstein tensor $\delta G_{AB} = \delta R_{AB} - \half g^{(0)}_{AB} \delta R$ have the form
\begin{align}
\label{linearEinstein}
\delta G_{rr}^\n &= -\frac{1}{2}\partial_r^2 g^\n_{ii},  \nonumber\\
\delta G_{ij}^\n &= -\frac{1}{2}\partial_r(r\partial_r g^\n_{ij}) - \frac{1}{2} \delta_{ij} \left( \partial_r^2 g^\n_{tt} - \partial_r(r\partial_r g^\n_{ij})\right ),  \nonumber\\
\delta G_{t i}^\n &= -r\delta G_{ri}^\n = -\frac{r}{2}\partial_r^2 g_{t i}^\n, \nonumber\\
\delta G_{t t}^\n &=-r \delta G_{r t}^\n = -\frac{r}{4}\left(2 r \partial_r^2 g^\n_{ii} + \partial_r g^\n_{ii}\right).
\end{align}
We define $g_{ii}^\n \equiv \delta^{ij} g_{ij}^\n$ and $\delta G_{ii}^\n \equiv \delta^{ij}\delta G_{ij}^\n$. In contrast, there is no change to the Lovelock tensor (\ref{Lovelock}) at the same order $n$ since the curvature of the Rindler background is zero and any term in the variation would contain some factor of curvature at zero order.

We want to find the $g^\n_{AB}$ that cancels out the $O(\epsilon^n)$ part of the field equations arising from the pre-existing solution at $O(\epsilon^{n-1})$. That is, we require
\beq
\delta G^\n_{AB} + \hat{G}^\n_{AB} + 2 \alpha \hat{H}^\n_{AB} = 0
\eeq
where the hat denotes the parts of the curvature arising from the pre-existing solution. In order for this set of equations to be consistent, one must impose the integrability conditions
\begin{align}
\hat{E}^\n_{tt} + r\hat{E}^\n_{tr} = 0 \label{int1}\\
\hat{E}^\n_{ti} + r\hat{E}^\n_{ri} = 0 \label{int2}\\
\partial_r(\hat{E}^\n_{tr}+ r \hat{E}^\n_{rr}) + (1/2) \hat{E}^\n_{rr} = 0
\end{align}
where we have defined $\hat{E}^\n_{AB} = \hat{G}^\n_{AB} + 2 \alpha \hat{H}^\n_{AB}$. These are consistent with the Bianchi identity and (\ref{int1}) follows from the conservation of the Brown-York stress tensor extended to Gauss-Bonnet gravity \cite{Davis:2002gn}, i.e.
\beq
(G_{A \nu} + 2 \alpha H_{A \nu}) N^A = \partial^\mu T_{\mu \nu} = 0, \label{conservation}
\eeq
where
\beq
T_{\mu \nu} =  2 (K \gamma_{\mu \nu} - K_{\mu \nu}) + 4 \alpha (J \gamma_{\mu \nu} - 3 J_{\mu \nu} - 2 \hat{P}_{\mu \rho \nu \sigma} K^{\rho \sigma}).  \label{stresstensor}
\eeq
The symbol $\hat{P}_{\mu \rho \nu \sigma} = \hat{R}_{\mu \rho \nu \sigma} + 2 \hat{R}_{\rho[\nu} \gamma_{\sigma] \mu}- 2 \hat{R}_{\mu [\nu} \gamma_{\sigma] \rho} + \hat{R} \gamma_{\mu [\nu} \gamma_{\sigma] \rho} $ is the divergence free part of the induced Riemann tensor and can be neglected here because we work with a flat induced metric, while
\beq
J_{\mu \nu} = \frac{1}{3} (2 K K_{\mu \sigma} K^\sigma_\nu + K_{\sigma \lambda} K^{\sigma \lambda} K_{\mu \nu} - 2 K_{\mu \sigma} K^{\sigma \lambda} K_{\lambda \nu} - K^2 K_{\mu \nu}).
\eeq

Using (\ref{linearEinstein}), one can solve the differential equations subject to two conditions: (i) that $g^\n_{AB} = 0$ at $r=r_c$ (the metric on $S_c$ remains flat) and (ii) that there is no singularity at $r=0$. The resulting solution is
\begin{align}
g^\n_{tt} = (1-r/r_c) F^\n_t(t,x^i) + \int^{r_c}_{r} dr' \int^{r_c}_{r'} dr'' \frac{2}{3} \left(\hat{E}^\n_{ii} - 4 \hat{E}^\n_{tr}- 2r\hat{E}^\n_{rr} \right) \label{gtt}\\
g^\n_{ti} = (1-r/r_c) F^\n_i(t,x^i) - 2 \int^{r_c}_{r} dr' \int^{r_c}_{r'} dr'' \hat{E}^\n_{ti} \label{gtr} \\
g^\n_{ij} = -2 \int^{r_c}_{r} dr' \frac{1}{r} \int^{r'}_{0} dr'' \left(\hat{R}_{ij}^\n + 2 \alpha (\hat{H}_{ij}^\n - \frac{1}{3} \hat{H}_{kk}^\n)  \right) \label{gij}
\end{align}
where $F^\n_t(t,x^i)$ and $F^\n_i(t,x^i)$ are arbitrary functions.

These two remaining functions can be fixed by imposing gauge choices on the Brown-York stress tensor of the fluid (\ref{stresstensor}). The addition of the new metric piece at $O(\epsilon^n)$ has the following effect on the extrinsic curvature at the same order
\beq
\delta K_{\mu \nu}^\n =\left. \half \sqrt{r_c} \partial_r g^\n_{\mu \nu}\right|_{S_c}
\eeq
implying that
\beq
\delta K_{tt}^\n = -\frac{F_t^\n(t,x^i)}{2\sqrt{r_c}}, \qquad \delta K_{t i}^\n = -\frac{F_i^\n(t,x^i)}{2\sqrt{r_c}},
\qquad \delta K^\n_{ij} =  +\frac{1}{\sqrt{r_c}}\int_0^{r_c} d r' \left( \hat{R}_{ij}^\n + 2 \alpha (\hat{H}_{ij}^\n - \frac{1}{3} \d_{ij} \hat{H}_{kk}^\n) \right).
\eeq
By explicit calculation, we verified that there is no corresponding $O(\epsilon^n)$ variation of the $J_{\mu \nu}$ part of the stress tensor. Thus, the variation $\delta T_{\mu \nu}^\n$ comes only from the linear part in the extrinsic curvature:
\begin{align}
\delta T_{tt}^\n = -\sqrt{r_c} \int^{r_c}_0 2 R_{ii}^\n, \qquad  \delta T_{ti}^\n = \frac{F^\n_i(t,x^i)}{\sqrt{c}} \nonumber \\[1ex]
\delta T_{ij}^\n = \delta_{ij} \left(\frac{F^\n_t(t,x^i)}{r_c^{3/2}} + \frac{2}{\sqrt{r_c}} \int^{r_c}_{0} dr' (R_{kk}^\n + \frac{2\alpha}{3} \hat{H}_{kk}^\n) \right) - \frac{2}{\sqrt{r_c}} \int^{r_c}_{0} dr' (\hat{R}_{ij}^\n + 2\alpha \hat{H}_{ij}^\n). \label{perturbedstress}
\end{align}
The complete stress-tensor has the form
\beq T_{\mu \nu}^\n = \delta T_{\mu \nu}^\n + 2 \left(\hat{K}^\n \gamma_{\mu \nu} - \hat{K}_{\mu \nu}^\n\right) + 4 \alpha \left(\hat{J}^\n \gamma_{\mu \nu} - 3 \hat{J}_{\mu \nu}^\n \right), \label{stressfull}
\eeq
where as before, the hat notation indicates the part of the stress-tensor originating from the solution at $O(\epsilon^{n-1})$. The function $F^\n_i(t,x^i)$ is fixed by imposing the Landau gauge condition (\ref{Landau}) order by order in the non-relativistic expansion. This plays a role only at odd orders in $\epsilon$. The other function $F^\n_t(t,x^i)$, which appears at even orders, is fixed by requiring that there are no higher order corrections to the definition of the non-relativistic pressure, i.e. the isotropic part of $T_{ij}$ is
\beq T^{iso}_{ij} = \left(\frac{1}{\sqrt{r_c}} + \frac{P}{r_c^{3/2}}\right) \delta_{ij} \label{pressuregauge}\eeq
at all orders.

\subsection{Solution to $O(\epsilon^5)$}

We now apply the algorithm to solve for the metric to $O(\epsilon^5)$. One first starts with the seed metric solution (\ref{seedmetric}) and constructs the solution at $O(\epsilon^3)$. As we argued earlier, the corrections due to the Gauss-Bonnet coupling constant arise at $O(\epsilon^4)$. Therefore, the Gauss-Bonnet terms do not contribute and the solution reduces to the GR one found previously in \cite{Compere:2011dx}, where the only non-vanishing component is
\begin{align}
g_{ti}^{(3)} = \frac{r-r_c}{2r_c} \left[\left(v^2+2P\right) \frac{2v_i}{r_c} + 4 \partial_i P - (r+r_c) \partial^2 v_i \right].
\end{align}

The next step is to compute the $\hat{R}_{AB}^{(4)}$ and $\hat{H}_{AB}^{(4)}$ using this metric. Via direct calculation of the Lovelock tensor (\ref{Lovelock}), we find that
\beq H_{ij}^{(4)} = \frac{3}{4 r_c^2} \left(\omega_{ik} \omega^{k}{}_j + \half \delta_{ij} \omega_{kl} \omega^{kl} \right) \label{H4}
\eeq
with all other components of $H_{AB}^{(4)}$ equal to zero. At even order in $\epsilon$, $R_{ti}= 0$ and as a result $g^{(4)}_{ti}=0$. The remaining components to compute are $R_{tt}^{(4)}$, $R_{rr}^{(4)}$, $R_{rt}^{(4)}$, and $R_{ij}^{(4)}$, which we will not display explicitly here.

Using (\ref{H4}), the solution for $g_{tt}^{(4)}$ in (\ref{gtt}) reduces to
\beq
g^{(4)}_{tt} = (1-r/r_c) F^{(4)}_t(t,x^i) + \int^{r_c}_{r} dr' \int^{r_c}_{r'} dr'' \left(\hat{R}^\n_{ii} + \frac{4}{3}\alpha \hat{H}^{(4)}_{ii}- 2 \hat{R}^{(4)}_{rt}-r \hat{R}^{(4)}_{rr}\right)
\eeq
and we find that
\begin{align}
g^{(4)}_{tt} = (1-r/r_c) F_t^{(4)}(t,x^i) + \frac{(r-r_c)^2}{8r_c} \left(8 v_k \partial^2 v^k - \sigma_{kl} \sigma^{kl} \right) + \frac{(r-r_c)^2 (r-r_c+2 \alpha)}{8 r_c} \omega_{kl} \omega^{kl}.
\end{align}
The gauge condition on the stress tensor (\ref{pressuregauge}) fixes
\begin{align}
F_t^{(4)}(t,x^i) = \frac{9}{8r_c}v^4+\frac{5}{2r_c}Pv^2+\frac{P^2}{r_c}-2r_c v_i \partial^2 v^i - \left(\frac{r_c+\alpha}{2}\right)\sigma_{kl}\sigma^{kl}- \frac{\alpha}{2} \omega_{kl} \omega^{kl} - 2\partial_t P + 2v^k \partial_k P.
\end{align}
Note that in these expressions we have imposed incompressibility $\partial_i v^i =0$, used the Navier-Stokes equation (\ref{NS}) to eliminate time derivatives of $v_i$, and imposed
\beq \partial^2 P = - \partial_i v_j \partial^j v^i, \eeq
which also follows from (the divergence of) Navier-Stokes.
Meanwhile Eqn. (\ref{gij}), yields
\begin{align}
g^{(4)}_{ij} = &(1-\frac{r}{r_c}) \left[ \frac{1}{r_c^2}v_i v_j (v^2+2P)+\frac{2}{r_c}v_{(i}\partial_{j)}P-4\partial_i\partial_j P -\frac{1}{2}\sigma_{ik}\sigma^{k}{}_{j}+\frac{r-5r_c+12\alpha}{4r_c}\omega_{ik}\omega^{k}{}_{j} \right.\nonumber\\
&\quad + \sigma_{k(i}\omega_{j)}{}^{k}-\frac{r+r_c}{r_c}v_{(i}\partial^2 v_{j)}+\frac{r+5r_c}{4}\partial^2\sigma_{ij} \nonumber \\
& \quad -\left.\frac{1}{r_c}v_{(i}\partial_{j)}v^2-\frac{1}{2r_c}\sigma_{ij}(v^2+2P) + \frac{\alpha}{r_c} \delta_{ij} \omega_{kl} \omega^{kl} \right]. \label{g42}
\end{align}

We now use (\ref{stressfull}) and (\ref{perturbedstress}) to find the stress tensor components $T^{(4)}_{tt}$ and $T^{(4)}_{ij}$. The non-zero components of the $J^{(4)}_{\mu \nu}$ tensor are
\begin{align}
J^{(4)}_{tt} = -\frac{1}{24 r_c^{1/2}} \sigma_{ij} \sigma^{ij}, \quad J^{(4)}_{ij} = \frac{1}{12 r_c^{3/2}} \sigma_{ik} \sigma^{k}{}_{j}
\end{align}
Using this result, we find
\begin{align}
T^{(4)}_{tt} = r_c^{-3/2}\Big[v^2(v^2+P)-\frac{r_c^2}{2}\sigma_{ij}\sigma^{ij}-r_c \sigma_{ij}v^i v^j\Big]
\end{align}
and
\begin{align}
T^{(4)}_{ij} =&  r_c^{-5/2}\left[v_i v_j (v^2+P) +2r_c v_{(i}\partial_{j)}P -4 r_c^2 \partial_i \partial_j P-\frac{r_c^2}{2}\left(1+\frac{2\alpha}{r_c}\right)\sigma_{ik} \sigma^{k}{}_{j}-r_c^2 \left(1+\frac{3\alpha}{r_c} \right)\omega_{ik} \omega^{k}{}_{j} \right.\nonumber\\[1ex]
&\quad + \left. r_c^2 \sigma_{k(i} \omega_{j)}{}^{k} -2 r_c^2 v_{(i}\partial^2 v_{j)}+\frac{3r_c^3}{2}\partial^2 \sigma_{ij}-r_c v_{(i}\partial_{j)}v^2 -\frac{r_c}{2}\sigma_{ij} v^2  \right].
\end{align}
Note that the $T^{(4)}_{tt}$ has no $\alpha$ corrections. They cancel out and the energy density $T_{\mu \nu} u^\mu u^\nu$ is not affected by $\alpha$ at fourth order. Comparing with the general form of the fluid stress tensor (\ref{T4}) we read off that
\begin{align}
b_1 = -\sqrt{r_c}\,,\: b_2 = 0\,,\: c_1=-2\sqrt{r_c}\left(1+\frac{2\alpha}{r_c}\right),\: c_3 = -4\sqrt{r_c}\left(1+\frac{3\alpha}{ r_c}\right), \: c_2=c_4=-4\sqrt{r_c}
\end{align}
as expected, there is no change in the value of $\eta=1$. However, the Gauss-Bonnet term does modify the two transport coefficients $c_1$ and $c_3$ from their purely GR values.

\section{Discussion}

We have argued that higher curvature corrections to the Einstein equations always come in at $O(\epsilon^4)$ in the non-relativistic hydrodynamic expansion and at $O(\lambda^2)$ in the relativistic Knudsen number expansion, at least when we perturb the fluid dual to the flat Rindler spacetime geometry. Hence, the solution to the higher curvature theories at the lowest orders is the same as the GR solution found previously \cite{Bredberg:2011jq, Compere:2011dx}.
Working in the specific case where the higher curvature theory is Einstein-Gauss-Bonnet gravity,  we then showed explicitly that the 1st order viscous hydrodynamics of the dual fluid is the same both in Einstein gravity and its higher curvature generalization, while the effect of the higher curvature corrections shows up in the second order transport coefficients of the fluid. We calculated some of these transport coefficients and found that two of them depend on the Gauss-Bonnet coupling constant. It would be interesting to complete the relativistic calculation outlined in Section III B in order to find all the second order transport coefficients in both the Einstein and Einstein-Gauss-Bonnet examples.

The approximate strong universality \cite{Coley:2008th} of the seed solution about which the hydrodynamic expansion is made is an interesting result. The lack of a higher curvature correction to the viscosity implies that it is protected against quantum corrections or other deformations to the dual theory. One way of thinking about these results is to note that shear viscosity and entropy density typically scale like $T^d$, where $T$ is the equilibrium temperature of the thermal system. In AdS/CFT, this temperature is given by the Hawking temperature $T_H$ of the black brane solution, which would depend in this case on the Gauss-Bonnet coupling constant, due to the non-trivial curvature of the background solution. In contrast, the shear viscosity and entropy density are constants independent of the temperature in the Rindler case.  This suggests the two types of holographic duality are different.

The independence of the fluid/Rindler holographic duality from the asymptotic geometry makes this correspondence interesting beyond the AdS/CFT context.  For example, the Rindler metric is associated to an accelerated observer in the locally flat surroundings of any point in spacetime. Therefore, one can ask whether the flat spacetime duality can be applied locally and then possibly used to patch together a holographic description of any spacetime \cite{Compere:2011dx}.

In a similar manner, the local Rindler system is also crucial the idea that gravity may emerge from the holographic hydrodynamics of some microscopic, quantum system \cite{Jacobson:1995ab,Eling:2006aw,Eling:2008af,Chirco:2009dc}. Here one assumes that the local Minkowski vacuum state carries a finite area entanglement entropy which can be holographically identified with the entropy of the local Rindler horizon.  Perturbations to the horizon system are assumed to obey an entropy balance law, relating a change in the entropy to the ``heat" associated with a flux of matter, plus an internal entropy production term from shear viscosity. Demanding that this equation holds at each point in spacetime then yields the GR Einstein equation and fixes the shear viscosity to entropy density ratio to be $1/4\pi$.  Also, bulk viscosity does not appear as an independent transport coefficient \cite{Eling:2008af, Chirco:2010sw}, which is strikingly similar to the viscous hydrodynamics of the global Rindler fluid.

However, extensions of this type of derivation to $f(R)$ gravities \cite{Eling:2006aw, Chirco:2009dc, Chirco:2010sw} require the horizon entropy to depend on the curvature, which inevitably leads to the same behavior in the shear viscosity. Ultimately, while the metric around any point is flat, the curvature itself does not vanish at the point. This fact means that one cannot simply import results pertaining to perturbations of the globally flat Rindler solution into the locally flat patch. On the other hand, there is some evidence that the inherent fuzziness in the local Killing vector, which is associated with the local notion of thermal equilbrium,  may be of the same order of magnitude as higher curvature corrections \cite{Eling:2006aw}. If this is the case, the approximate notion of a local fluid would not be affected by these corrections. It would be interesting to investigate further the relationship between the fluid/Rindler correspondence and these ideas of emergent  graviational dynamics.


Finally, to conclude, we want to point out an interesting duality here between the relativistic $\lambda$ expansion in derivatives in the holographic theory and an effective field theory expansion of the bulk gravitational theory. First note that the $\lambda$ expansion is equivalent to an expansion in small dimensionless \textit{Knudsen number}, which is defined as $Kn =\frac{\ell_{\rm{mfp}}}{L}$, where $\ell_{\rm{mfp}}$ is the mean free path associated with the microscopic system and $L$ is the characteristic size of the perturbations to the system. Secondly, although the bulk gravity theory is non-renormalizable, it is still valid as an effective theory when the dimensionless ratio of the Planck length to the radius of curvature, $\frac{L_{\rm{planck}}}{R_{\rm{curv}}}$, is small. The effective action is given as an expansion in this ratio. At zeroth order there is some cosmological constant, at first order, the Hilbert term, and then the pieces higher order in curvature invariants.

In the duality, the scale of perturbations $L$ in the system on $r=r_c$ is linked to the scale $R_{\rm{curv}}$ of perturbations to the flat bulk spacetime. As we have seen, a flat spacetime is dual to the fluid in equilibrium, Einstein gravity dual to the viscous hydrodynamics characterized by a shear viscosity, and second order transport coefficients linked to curvature squared terms. It is tempting to associate the universality of the shear viscosity with the universality of the Hilbert action at low energies and take $\ell_{\rm{mfp}} \sim L_{\rm{planck}}$, the scale at which gravity is strongly coupled. This line of reasoning also suggests it may be interesting to consider a seed metric constructed from the region of a de Sitter spacetime where there is also a causal ``observer dependent" horizon and the associated thermodynamics. What effect does a non-zero cosmological constant have on the dual fluid?

\section*{Acknowledgements}

We thank M. Cadoni and Y. Oz for valuable discussions and P. McFadden for a clarification.

\end{document}